\documentclass[a4paper,11pt]{article}
\usepackage[dvips]{graphicx}
\usepackage{subeqn}
\begin{document}
\baselineskip 0.7cm
\def\ie{{\it i.e.}}
\newcommand{\gsim}{ \mathop{}_{\textstyle \sim}^{\textstyle >} } 
\newcommand{\lsim}{ \mathop{}_{\textstyle \sim}^{\textstyle <} } 
\def\mMEt{\not\kern-.35em {E_T}}
\def\MEt{\hbox{$\mMEt$}}

%--------------------
% \slashchar puts a slash through a character to represent contraction
% with Dirac matrices. Use \not instead for negation of relations, and use
% \hbar for hbar.
\def\slashchar#1{\setbox0=\hbox{$#1$} % set a box for #1
\dimen0=\wd0 % and get its size
\setbox1=\hbox{/} \dimen1=\wd1 % get size of /
\ifdim\dimen0>\dimen1 % #1 is bigger
\rlap{\hbox to \dimen0{\hfil/\hfil}} % so center / in box#1 
% and print #1
\else % / is bigger
\rlap{\hbox to \dimen1{\hfil$#1$\hfil}} % so center #1
/ % and print /
\fi}
%--------------------
\setcounter{footnote}{1}
%%%%%%%%%%%%%%%%%%%%%%%%%%%%%%%%%%%%%%%%%%%%%%
%%%%%%%                                %%%%%%%
%%%%%%%	        Cover page             %%%%%%%
%%%%%%%                                %%%%%%%
%%%%%%%%%%%%%%%%%%%%%%%%%%%%%%%%%%%%%%%%%%%%%%
\thispagestyle{empty}
\vspace*{-15mm}
%----------
\baselineskip 10pt
\begin{flushright}
\begin{tabular}{l}
{\bf OCHA-PP-267}\\
{\bf hep-ph/0611240}
\end{tabular}
\end{flushright}
\baselineskip 18pt 
\vglue 10mm 
%%%%%%%%%%%%%%%%%%%%%%%%%%%%%%%%%%%%%%%%%%%%%%
%                Title 
%%%%%%%%%%%%%%%%%%%%%%%%%%%%%%%%%%%%%%%%%%%%%%
\begin{center}
{\Large\bf
%Decay of Charged Higgs boson in TeV scale supersymmetric 
%seesaw model
%
Decay of Charged Higgs boson in a scenario of SUSY breaking 
inspired neutrino mass
}
\vspace{13mm}

\baselineskip 18pt 
\def\thefootnote{\fnsymbol{footnote}}
\setcounter{footnote}{0}
{\bf
Gi-Chol Cho$^{a)}$, Satoru Kaneko$^{a)}$\footnote{
present address: AHEP Group, Instituto de F\'{\i}sica Corpuscular -- 
C.S.I.C./Universitat de Val{\`e}ncia, Edifici Instituts d'Investigacio, 
Apt. 22085, E--46071 Val{\`e}ncia, Spain} 
and Aya Omote$^{b)}$ 
}

\vspace{10mm}

$^{a)}$ {\it Department of Physics, Ochanomizu University, 
Tokyo 112-8610, Japan}
\\
$^{b)}$ {\it Graduate School of Humanities and Sciences, 
Ochanomizu University, Tokyo, 112-8610, Japan}
\vspace{10mm}
\end{center}
%%%%%%%%%%%%%%%%%%%%%%%%%%%%%%%%%%%%%%%%
%%%%%                              %%%%%
%%%%%          Abstract            %%%%%
%%%%%                              %%%%%
%%%%%%%%%%%%%%%%%%%%%%%%%%%%%%%%%%%%%%%%
\begin{center}
{\bf Abstract}\\[7mm]
\begin{minipage}{16cm}

In some class of supersymmetric models, small neutrino mass is given as 
a consequence of the supersymmetry (SUSY) breaking. 
Phenomenologically interesting features of this scenario are as follows: 
(i) the right-handed sneutrino mass could be as low as TeV scale due to 
the Giudice-Masiero mechanism, and (ii) a scalar trilinear 
interaction of Higgs-slepton-(right-handed) sneutrino could be sizable 
without suppression by the small neutrino Yukawa coupling. 
We study some phenomenological aspects of this scenario focusing on 
the scalar trilinear interaction. 
We show that the 1-loop correction by sneutrino exchange to the lightest 
Higgs boson mass destructively interferes with top-stop contributions in 
the minimal SUSY Standard Model.  
We find that a decay of charged Higgs boson into sneutrino and charged 
slepton is sizably enhanced and hence it gives rise to a distinctive 
signal at future collider experiments in some parameter space. 
\baselineskip 16pt
\noindent
%%%%%----------------------------------   
\end{minipage}
\end{center}
%%%%%------------------------------------------
%%%%% PACS number(s) & Key words
%%%%%------------------------------------------
%%%\baselineskip 18pt 
%%%{\small 
%%%\begin{flushleft}
%%%{\sl PACS}: 11.30.Pb, 12.15.Lk, 12.60.Jv\\
%%%{\sl Keywords}: 
%%%\end{flushleft}
%%%}
%%%%%---------------------------------- 
\newpage
%%%%%%%%%%%%%%%%%%%%%%%%%%%%%%%%%%%%%%%%%%%%%%%%%%%%%%%%%%
\newpage
%\baselineskip=20 pt
%%%%%%%%%%%%%%%%%%%%%%%
\section{Introduction}
%%%%%%%%%%%%%%%%%%%%%%%
Smallness of neutrino mass is one of the important clues to physics 
beyond the Standard Model (SM). 
An attractive explanation on the origin of small neutrino mass is 
the seesaw mechanism~\cite{seesaw}. 
In the seesaw mechanism, a heavy right-handed neutrino is introduced 
and it couples to SU(2)$_L$ doublet neutrino and Higgs boson through 
the Yukawa coupling $Y_\nu$. 
After diagonalizing the neutrino mass matrix, 
a smaller mass eigenvalue of neutrino, $m_\nu$, is given by 
\begin{eqnarray}
m_\nu\simeq (Y_\nu v)^2/m_N, 
\label{seesaw}
\end{eqnarray}
where $m_N$ and $v$ are the mass of 
right-handed neutrino and the vacuum expectation value (v.e.v.) of the 
Higgs boson, respectively. 
If the Yukawa coupling $Y_\nu$ is of order unity, the right-handed 
neutrino should be heavy enough, say, 
$m_N \sim 10^{11}{\rm GeV}/(m_\nu/1{\rm eV})$, to be consistent with
results of neutrino experiments. 
Then one may complain the large hierarchy between the scale of the 
seesaw mechanism ($m_N$) and the electroweak scale ($v$). 
Furthermore it is hopeless to confirm the seesaw mechanism through 
searching for the right-handed neutrino at collider experiments 
(a trial to test the seesaw mechanism with hypothetical outcome 
of future experiments is proposed in ref.\cite{Buckley:2006nv}). 
Thus it may be worth considering a possibility to lower the scale 
of seesaw mechanism (scale of right-handed neutrino) as low as 
testable at collider experiments, say, $O(100{\rm GeV}-1{\rm TeV})$, 
or alternative to the seesaw mechanism from a phenomenological point of view. 
%%% 

%%%
It has been argued possibilities to explain the small neutrino 
mass as a consequence of supersymmetry (SUSY) breaking 
in refs.~\cite{Mur,Borzumati:2000mc,March-Russell:2004uf}. 
Some phenomenologically viable points of this class of models 
are (i) light (TeV scale) right-handed sneutrino due to the 
Giudice-Masiero mechanism~\cite{GM} and (ii) enhancement of 
scalar trilinear interaction among the right-handed sneutrino, 
left-handed slepton and Higgs bosons. 
Both (i) and (ii) can be, for example, realized as follows. 
Let us first introduce a chiral superfield $X$ which is a 
SM gauge singlet but charged under a certain global symmetry. 
This global symmetry may allow non-renormalizable operators 
such as 
\begin{eqnarray}
 \frac{X X^\dagger}{M_P^2} N^\dagger N, 
\label{mr}
\end{eqnarray}
and 
\begin{eqnarray}
 \frac{X}{M_P} L H_u N, 
\label{aterm}
\end{eqnarray}
where dimensionless couplings with $O(1)$ magnitude are suppressed. 
In (\ref{mr}) and (\ref{aterm}), 
$L$ and $N$ denote the left-handed lepton and right-handed 
neutrino superfields, respectively. 
The Higgs superfield with the hypercharge $Y=1/2$ is represented 
by $H_u$, and $M_P$ is the reduced Planck mass. 
Suppose that the $F$-component of the $X$ field develops a v.e.v. 
$\langle F \rangle \sim m_{3/2} M_P$ due to the SUSY breaking, 
where $m_{3/2}$ is the gravitino mass. 
Then the $D$-component of (\ref{mr}) leads to the right-handed sneutrino 
mass as 
\begin{eqnarray}
\left( \frac{X X^\dagger}{M_P^2} N^\dagger N\right)_D 
\to m^2_{\widetilde{\nu}_R} 
\widetilde{\nu}_R^* \widetilde{\nu}_R, 
\end{eqnarray}
while the $F$-component of (\ref{aterm}) gives the scalar trilinear 
interaction 
\begin{eqnarray}
\left( \frac{X}{M_P} L H_u N \right)_F 
\to A_\nu \widetilde{\ell} H_u \widetilde{\nu}_R. 
\end{eqnarray}
Note that both $m_{\widetilde{\nu}_R}$ and $A_\nu$ are 
of order the gravitino mass $\sim O({\rm TeV})$. 
Moreover the scalar trilinear interaction is not suppressed 
if a dimensionless coupling in (\ref{aterm}) is of order unity. 
%%%

%%% 
In the minial SUSY SM (MSSM), the SUSY breaking scalar trilinear 
interactions of squark or sleptons are parametrized by $A_f Y_f$, 
where $A_f$ and $Y_f$ are the scalar trilinear 
coupling and the Yukawa coupling for flavor $f$, respectively. 
The scalar three-point vertices are, therefore, suppressed by small 
Yukawa couplings for the first two generations of squarks and 
sleptons. 
In the models of 
refs.~\cite{Mur,Borzumati:2000mc,March-Russell:2004uf}, 
however, the scalar trilinear interaction of the right-handed sneutrino 
is not suppressed by the neutrino Yukawa coupling, as mentioned above. 
%%%

%%% 
A few comments on this class of models are in order. 
In a series of non-renormalizable operators ((\ref{mr}), 
(\ref{aterm}), etc), there are lepton-number violating operators 
in general. 
If such operators are forbidden by an appropriate discrete symmetry, the 
seesaw mechanism does not work and the Dirac neutrino mass 
should be given by (\ref{aterm}) with the $A$-component v.e.v. 
$\langle A \rangle$ of the $X$-field. 
Then, to satisfy the experimental limit, a relation 
$\langle A \rangle \ll \sqrt{\langle F \rangle}$ must hold. 
On the other hand, if the lepton-number violating operators are allowed, 
the Majorana mass term of the right-handed neutrino and the SUSY breaking 
$B$-term of right-handed sneutrino appear at the same order of magnitude. 
This $B$-term contributes to the smaller neutrino mass through 
the radiative correction and, therefore, should be highly suppressed. 
Further discussions on these constraints can be found in 
refs.~\cite{Borzumati:2000mc,BHNY}. 
%%%

%%% 
In this paper, we investigate phenomenological consequences of 
a scenario of TeV scale right-handed sneutrino inspired by 
supersymmetric models in 
refs.~\cite{Mur,Borzumati:2000mc,March-Russell:2004uf}, 
focusing on the unsuppressed coupling $A_\nu$. 
We first study the 1-loop corrections to the lightest Higgs boson 
mass through the sneutrino exchange which is proportional to some powers
of $A_\nu$. 
We show that the sneutrino contribution to the lightest Higgs 
boson mass destructively interferes with the MSSM contribution. 
We next study decay processes of charged Higgs boson~\cite{Mur}. 
Owing to $A_\nu$, the decay of charged Higgs boson into 
the sneutrino and selectron could be enhanced as compared to the MSSM. 
We find that, in some parameter space, the branching ratio of this 
decay mode can be as large as $10\%$, and it may be detectable at future 
linear collider experiments. 
In our study, we neglect the generation mixing in both the left- and 
right-handed sneutrinos for simplicity. 
Although this scenario has a possibility if the neutrino is Majorana or 
Dirac, our study is available in both cases if the SUSY breaking 
$B$-term of sneutrino in the Majorana case is assumed to be small enough 
so that, in addition to suppress the 1-loop correction to the mass of 
lighter neutrino, the sneutrino mass matrix has common structure in both 
cases.  

%%%%%%%%%%%%%%%%%%%%%%% 
\section{Mass and Interactions}
%%%%%%%%%%%%%%%%%%%%%%% 
We first review the sneutrino masses and interactions to fix our 
notation. 
When the SUSY breaking $B$-term of sneutrino is neglected, the mass 
matrix of sneutrinos in a basis of 
($\widetilde{\nu}_L, \widetilde{\nu}_R$) is given by  
\begin{eqnarray}
M_{\tilde{\nu}}^2
&=&
\left(
\begin{array}{ccc}
 m^2_{\widetilde{\nu}_L} & A_\nu v \sin\beta     \\
 A_\nu v \sin\beta & m^2_{\widetilde{\nu}_R}
\end{array}
\right)
\ ,
\label{massmatrix}
\\
 m^2_{\widetilde{\nu}_L} &=&
m_L^2 + \frac{1}{2}\cos2\beta m_Z^2, 
\label{snleft}
\end{eqnarray}
where $m_L$ is the soft scalar mass for the SU(2)$_L$ doublet slepton 
while $m_{\widetilde{\nu}_R}$ is for the right-handed sneutrino. 
The angle $\beta$ is defined as $\tan\beta \equiv v_u/v_d$, where 
$v_u$ and $v_d$ are v.e.v. of the Higgs bosons with $Y=1/2$ and $-1/2$,
respectively. 
A parameter $v$ is normalized 
as $v\equiv \sqrt{v_u^2+v_d^2} \approx 246{\rm GeV}$. 
The mass matrix (\ref{massmatrix}) can be diagonalized using an 
unitary matrix $U_{\widetilde{\nu}}$: 
\begin{eqnarray}
\left(U_{\widetilde{\nu}}\right)^\dagger
M^2_{\widetilde{\nu}}
U_{\widetilde{\nu}}
={\rm diag}(m_{\widetilde{\nu}_1}, m_{\widetilde{\nu}_2}), 
~~~(m_{\widetilde{\nu}_1} < m_{\widetilde{\nu}_2}). 
\label{snmass}
\end{eqnarray}
In the MSSM, the sneutrino mass is given by (\ref{snleft}). 
Note that $m^2_{\widetilde{\nu}_L}$ (\ref{snleft}) 
satisfies the following relation with the mass of 
left-handed selectron $\widetilde{e}_L$ due to the SU(2)$_L$ symmetry: 
\begin{eqnarray}
m_{\tilde{e}_L}^2- m_{\tilde{\nu}_L}^2&=& (-1 + s_W^2)m_Z^2\cos2\beta. 
\label{mass}
\end{eqnarray}
Since $\cos2\beta<1$ for $\tan\beta>1$, the mass of sneutrino in the 
MSSM is always smaller than the selectron mass when $\tan\beta > 1$. 
On the other hand, the lighter sneutrino mass 
(\ref{snmass}) is independent of the selectron mass and can be much 
lighter than the sneutrino in the MSSM. 
%%%--------------------------------
%%% Fig. 0
%%%--------------------------------
\begin{figure}[t]
\begin{center}
\includegraphics[width=6.5cm,clip]{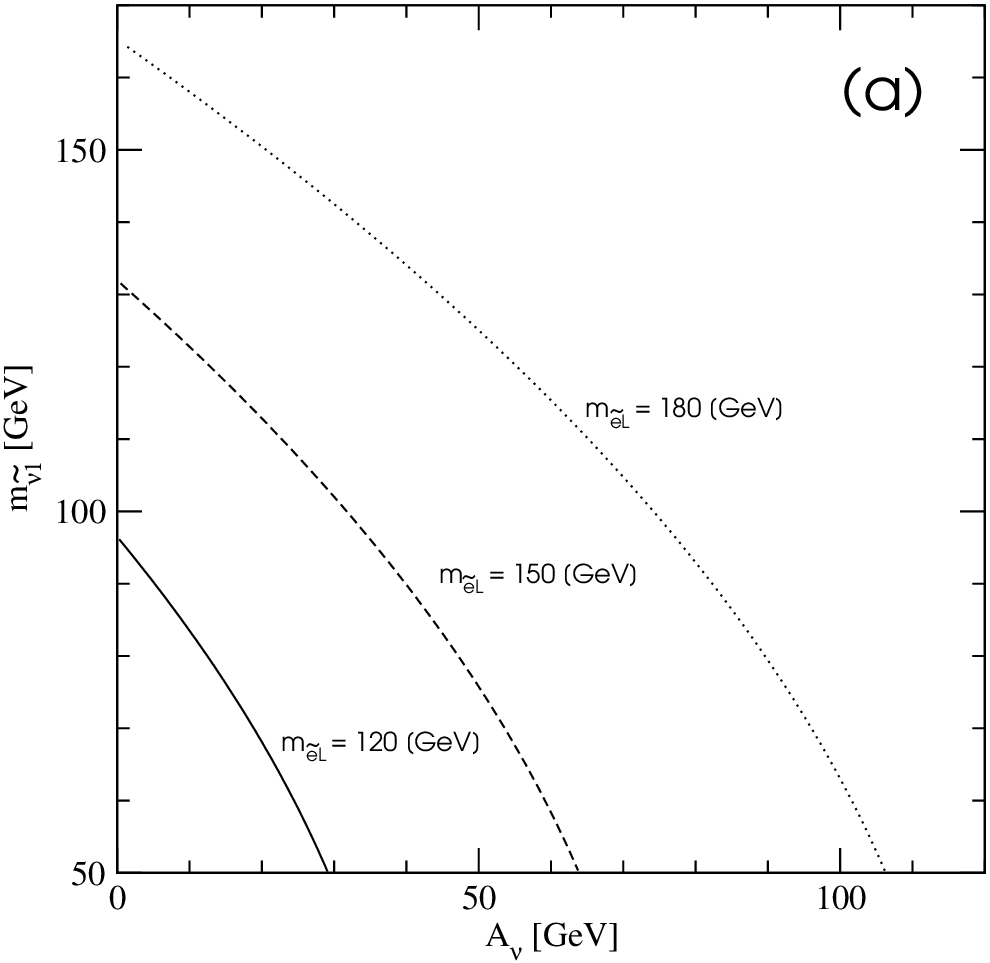}
\includegraphics[width=6.5cm,clip]{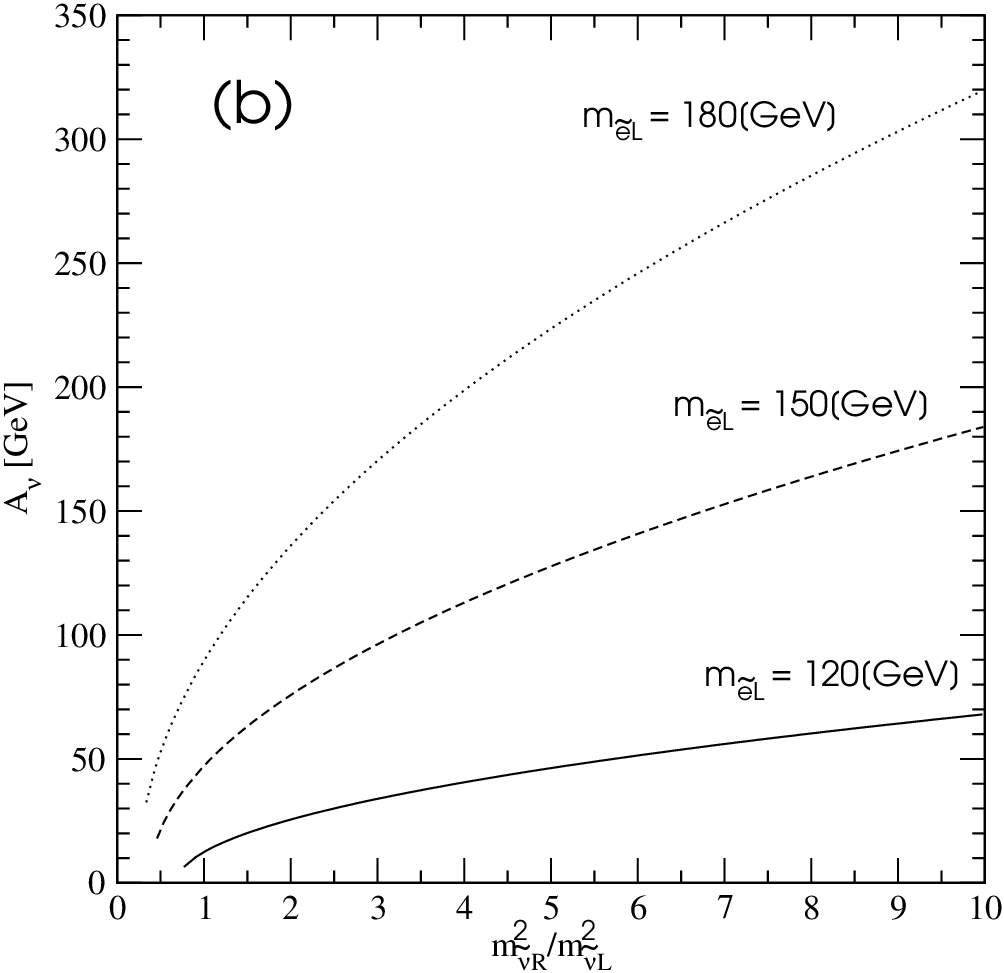}
\end{center}
\caption{
(a): The lighter sneutrino mass $m_{\widetilde{\nu}_1}$ as 
a function of $A_\nu$ for $\tan\beta=3$. 
Three lines correspond to $m_{\widetilde{e}_L}=120{\rm GeV}$ 
(solid), $150{\rm GeV}$ (dashed) and $180{\rm GeV}$ (dotted). 
The results are obtained by taking 
$m_{\widetilde{\nu}_L}^2=m_{\widetilde{\nu}_R}^2$. 
(b): The coupling $A_\nu$ as a function of 
a ratio $m_{\widetilde{\nu}_R}^2/m_{\widetilde{\nu}_L}^2$ 
for $\tan\beta=3$. 
}
\label{aandm}
\end{figure}
%%%

In Fig.~\ref{aandm}(a), we show the lighter sneutrino mass 
$m_{\widetilde{\nu}_1}$ as a function of $A_\nu$ for $\tan\beta=3$. 
Three lines correspond to $m_{\widetilde{e}_L}=120{\rm GeV}$ (solid), 
$150{\rm GeV}$ (dashed) and $180{\rm GeV}$ (dotted). 
For the right-handed sneutrino mass, we take 
$m_{\widetilde{\nu}_R}=m_{\widetilde{\nu}_L}$ for convenience. 
Note that the mass $m_{\widetilde{\nu}_1}$ at $A_\nu = 0$ corresponds 
to that in the MSSM. 
The figure tells us that the large left-right mixing of sneutrino which 
is induced by large $A_\nu$, makes a sneutrino much lighter than that in 
the MSSM.  
%%%

%%%
Fig.~\ref{aandm}(b) shows the trilinear coupling $A_\nu$ as a function 
of $m_{\widetilde{\nu}_R}^2/m_{\widetilde{\nu}_L}^2$. 
Three lines correspond to different values of the selectron 
mass as Fig.~\ref{aandm}(a).  
The lighter sneutrino mass $m_{\widetilde{\nu}_1}$ is fixed at 
$80{\rm GeV}$. 
It can be seen from the figure that the coupling $A_\nu$ increases when 
$m_{\widetilde{\nu}_R}^2/m_{\widetilde{\nu}_L}^2$ is larger than one.  
%%%

%%%
Next we summarize the interaction Lagrangian of sneutrino, slepton and 
Higgs bosons. 
For simplicity we take a limit of large pseudo scalar mass $m_A$. 
Then the lightest Higgs boson $h$ can be approximately identified with 
the SM Higgs.  
The interaction Lagrangians of sneutrino-sneutrino-lightest Higgs 
boson ($\widetilde{\nu}_i-\widetilde{\nu}_j-h$) 
and sneutrino-slepton-charged Higgs boson 
($\widetilde{\nu}_i-\widetilde{\ell}-H^-$) 
are then given as follows: 
\begin{itemize}
\item $\widetilde{\nu}_i-\widetilde{\nu}_j-h$ interaction: 
\begin{eqnarray}
{\cal L} &=& 
A_\nu 
\sum_{i,j}(U_{\widetilde{\nu}})^*_{1i}(U_{\widetilde{\nu}})_{2j}
\widetilde{\nu}_i^*\widetilde{\nu}_j h 
+ 
{\rm h.c.}~~~(i,j=1,2),
\label{intnh}
\end{eqnarray}
\item $\widetilde{\nu}_i-\widetilde{\ell}-H^-$ interaction: 
\begin{eqnarray}
{\cal L} &=& 
g^{\widetilde{\nu}_i \widetilde{\ell}H^-} \widetilde{\nu}_i^* 
\widetilde{\ell}H^{+}+{\rm h.c.},
\\
g^{\widetilde{\nu}_i \widetilde{\ell}H^-}
&=&
A_\nu \cos\beta (U_{\widetilde{\nu}})^*_{2i} 
-\frac{g}{\sqrt{2}}m_W\sin2\beta 
\left( U_{\widetilde{\nu}}\right)^*_{1i}
~~~(i=1,2). 
\label{intch}
\end{eqnarray}
\end{itemize}

%%%--------------------------------
%%% Fig. 1
%%%--------------------------------
\begin{figure}[t]
\label{higgs}
\begin{center}
\includegraphics[width=6.5cm,clip]{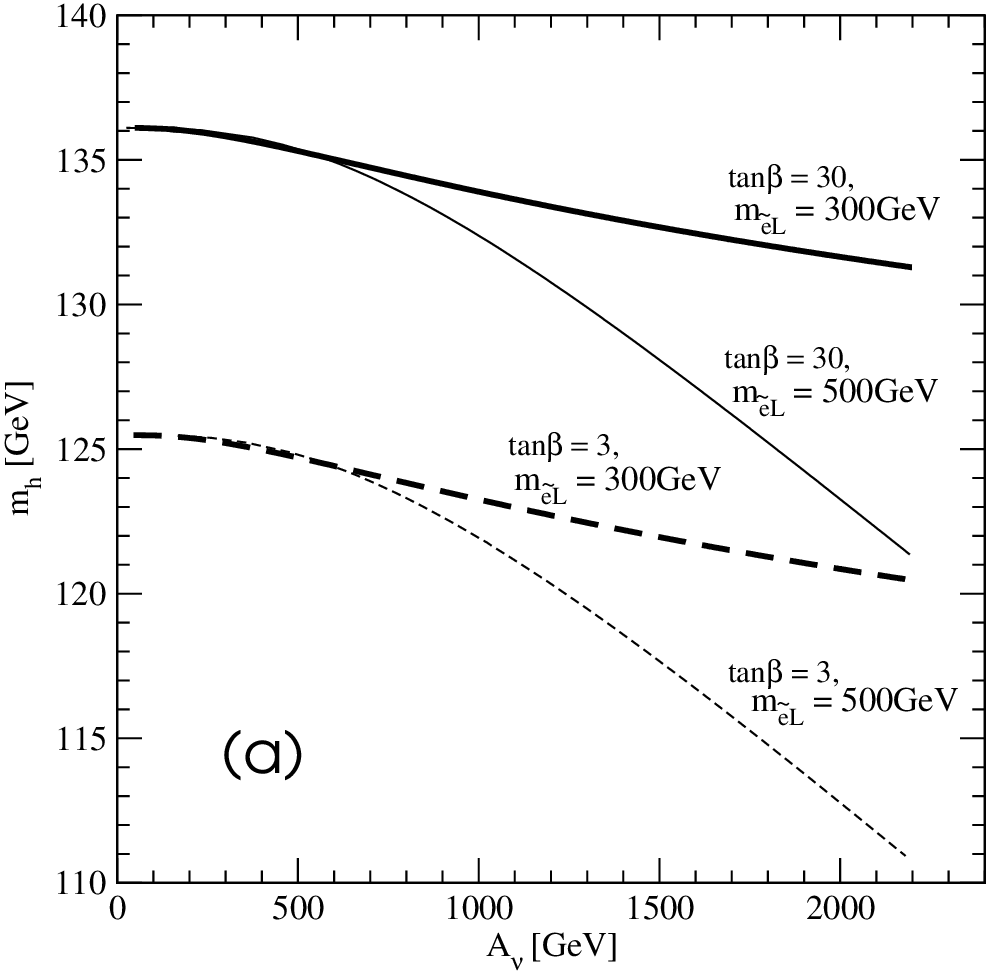}
\includegraphics[width=6.5cm,clip]{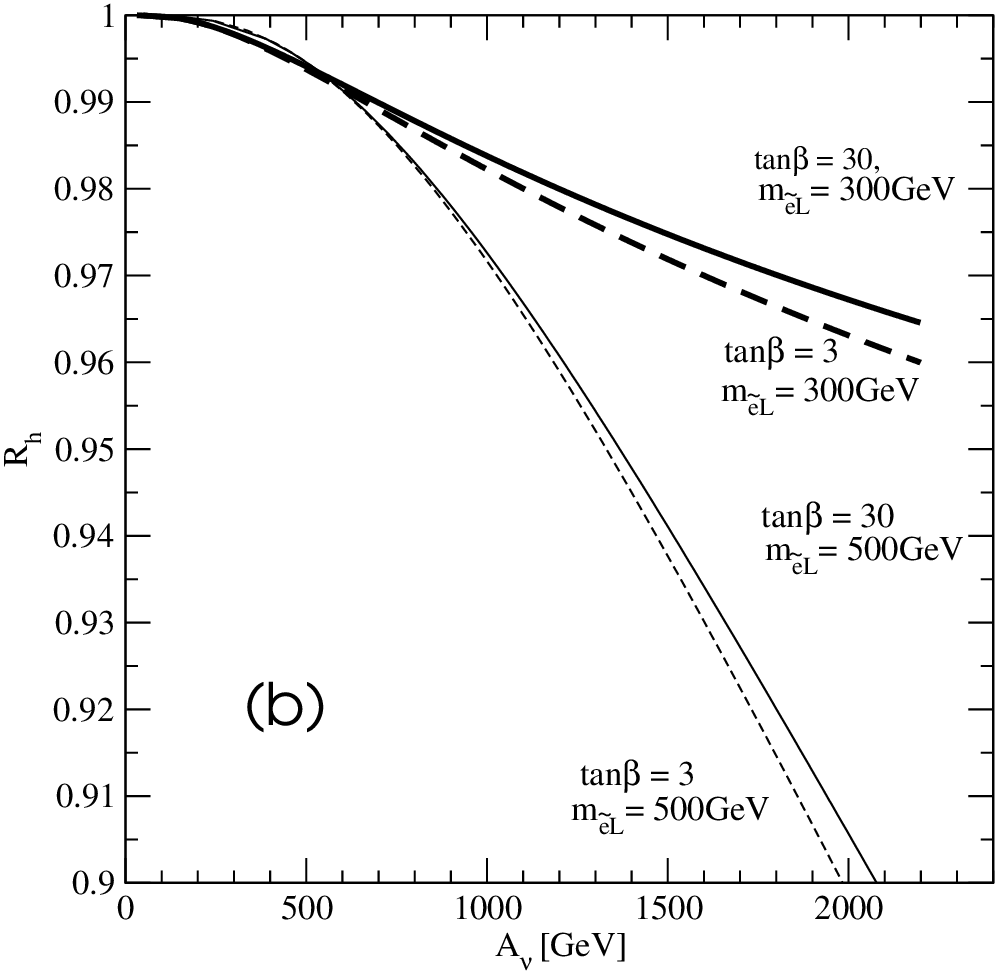}
\end{center}
\caption{ 
(a) 
The lightest Higgs boson mass $m_h$ as a function of $A_\nu$. 
Each line correspond to combinations of $\tan\beta=3,30$ 
and $m_{\widetilde{e}_L}=300,500{\rm GeV}$ as indicated. 
The 1-loop correction from the top-stop loop is evaluated 
following ref.~\cite{Haber:1996fp} using the stop mass 
$m_{\widetilde{t}}=1{\rm TeV}$. 
The Higgs mass $m_h$ at $A_\nu=0$ corresponds to the MSSM 
prediction. 
(b) 
The ratio $R_h$ defined in eq.(\ref{rh}) as a function of $A_\nu$. 
}
\end{figure}

%%%%%%%%%%%%%%%%%%%%%%%
\section{Sneutrino contribution to the lightest Higgs boson mass}

%%%%%%%%%%%%%%%%%%%%%%%
It is known that the lightest Higgs boson mass $m_h$ receives 
large 1-loop corrections mainly from the top quark and the 
stop exchanging diagram~\cite{Okada:1990vk,Okada:1990gg,Ellis:1990nz}. 
In the scenario of TeV scale $\widetilde{\nu}_R$ with sizable $A_\nu$, 
the $\widetilde{\nu}_L$-$\widetilde{\nu}_R$-$h$ 
interaction (\ref{intnh}) could give a new contribution to the 
lightest Higgs boson mass at 1-loop level.  
Using the renormalization group method used in 
ref.~\cite{Okada:1990gg}, we evaluate the sneutrino contribution to 
$m_h$. 
%%%

%%% 
Let us take the large limit of the SUSY breaking mass scale 
$m_{\rm SUSY}$ so that physics below $m_{\rm SUSY}$ is 
described by the Standard Model. 
Then the lightest Higgs boson mass $m_h$ is simply parametrized by 
\begin{eqnarray}
 m_h^2 &=& \lambda v^2,  
\end{eqnarray}
where $\lambda$ is a quartic coupling in the Higgs potential.  
Note that the quartic coupling at the tree level, $\lambda_{\rm tree}$, 
satisfies the SUSY relation 
\begin{eqnarray}
 \lambda_{\rm tree} = \frac{1}{4}(g_Y^2+g^2) \cos^2 2\beta, 
\end{eqnarray}
where $g_Y$ and $g$ are the U(1)$_Y$ and SU(2)$_L$ gauge 
couplings, respectively. 
The radiative corrections to the quartic coupling $\lambda$ 
in the MSSM can be found in, for example, ref.~\cite{Okada:1990gg}.  
In the scenario of large $A_\nu$, the interaction (\ref{intnh}) gives 
rise to the sneutrino exchanging box diagram as the 1-loop 
correction to the quartic coupling $\lambda$. 
The sneutrino contribution, $\lambda_{\widetilde{\nu}}$, 
can be evaluated as 
\begin{eqnarray}
\lambda_{\widetilde{\nu}} &=&
-\frac{A_\nu^4 }{4(4\pi)^2}
\sum_{i,j,k,l=1}^2
|U^{\widetilde{\nu}}_{1i}|^2 
|U^{\widetilde{\nu}}_{2j}|^2 
|U^{\widetilde{\nu}}_{1k}|^2 
|U^{\widetilde{\nu}}_{2l}|^2 
D_0(m_{\widetilde{\nu_i}},m_{\widetilde{\nu_j}},
m_{\widetilde{\nu_k}},m_{\widetilde{\nu_l}}), 
\end{eqnarray}
where the 1-loop scalar function $D_0$ is given by
\begin{eqnarray}
D_0(m_{1},m_{2},m_{3},m_{4})
&=&
\frac{1}{(m_1^2-m_2^2)(m_3^2-m_4^2)}
\left[
\frac{m_1^2+m_3^2}{m_1^2-m_3^2}\ln\frac{m_3}{m_1}
+\frac{m_2^2+m_4^2}{m_2^2-m_4^2}\ln\frac{m_4}{m_2}
\right.
\nonumber\\
&&
\left.
-\frac{m_1^2+m_4^2}{m_1^2-m_4^2}\ln\frac{m_4}{m_1}
-\frac{m_2^2+m_3^2}{m_2^2-m_3^2}\ln\frac{m_3}{m_2}
\right]. 
\end{eqnarray}
%%%

%%%
In Fig.~\ref{higgs}(a), we depict the $A_\nu$ dependence of the 
lightest Higgs boson mass $m_h$. 
We also compare, in Fig.~\ref{higgs}(b), a ratio of the Higgs boson 
mass in our scenario and in the MSSM which is defined as 
\begin{eqnarray}
R_h \equiv \frac{m_h}{m_h({\rm MSSM})}. 
\label{rh}
\end{eqnarray}
In eq.(\ref{rh}), $m_h$ and $m_h(\rm MSSM)$ are the lightest Higgs 
boson mass in our scenario and the Higgs mass in the MSSM, 
respectively. 
In the figures, the 1-loop corrections in the MSSM 
are estimated following ref.~\cite{Haber:1996fp} with the stop 
mass $m_{\widetilde{t}}=1{\rm TeV}$. 
Solid and dotted lines denote $m_{\widetilde{e}_L}=300{\rm GeV}$ 
and $500{\rm GeV}$, respectively. 
Thin and thick lines are $\tan\beta=3$ and 50 as indicated in the 
figures. 
Note that, in both figures, $m_h$ and $R_h$ at $A_\nu=0$ correspond to 
the MSSM prediction. 
It is easy to see that both $m_h$ and $R_h$ decrease when $A_\nu$ 
increases. 
This means that the sneutrino contribution to $m_h$ interferes 
with the MSSM contributions destructively. 
For example, in a limit where two sneutrino masses are equal 
($m_{\rm SUSY}$), the quartic coupling $\lambda_\nu$ is given as 
\begin{eqnarray}
\lambda_\nu \simeq 
 -\frac{1}{6}\frac{1}{(4\pi)^2}
\left(\frac{A_\nu}{m_{\rm SUSY}} \right)^4< 0. 
\label{coup}
\end{eqnarray} 
The minus sign in r.h.s. of (\ref{coup}) is the origin that $m_h$ is 
lowered via the sneutrino contribution.  
%
%The coupling $\lambda_\nu$ is roughly $~10^{-3}$ for 
%$A_\nu/m_{\rm SUSY} \sim O(1)$, but it could be sizable when $A_\nu > 
%m_{\rm SUSY}$ due to the fourth power of the ratio $A_\nu/m_{\rm SUSY}$. 
%
Fig.~\ref{higgs}(b) shows that the negative contribution 
to $m_h$ from the sneutrino diagram is less than $5\%$ for 
$A_\nu \lsim 1{\rm TeV}$. 

%%%%%%%%%%%%%%%%%%%%%%% 
\section{Decay of charged Higgs boson} 
\label{hd}
%%%%%%%%%%%%%%%%%%%%%%%
Next we examine a decay 
$H^- \to \widetilde{\nu} + \widetilde{\ell}$, where $H^-$ stands for 
a charged Higgs boson. 
In particular, a case of $\widetilde{\ell}=\widetilde{e}$ could be 
a distinctive process of our scenario because that such 
process is strongly suppressed in the MSSM due to the electron 
Yukawa coupling. 
So, we consider only the case of $\widetilde{\ell}=\widetilde{e}$ 
in the following study. 
In the MSSM, it is known that, for $m_{H^-} \gsim 200{\rm GeV}$, 
$H^-$ dominantly decays into the top and bottom quarks owing to 
the sizable Yukawa couplings (for a review of various decay 
channels of the charged Higgs boson in the supersymmetric 
models, see ref.~\cite{higgs}). 
The $\tau+ \nu_{\tau}$ mode is subdominant for large 
$\tan\beta(\gsim 10)$ due to the tau-Yukawa coupling. 
On the other hand, when $A_\nu$ is sizable, it is expected that 
the decay mode $H^- \to \widetilde{\nu}_1 + \widetilde{e}$ 
is much enhanced in small $\tan\beta$ region because that 
the decay vertex is proportional to $A_\nu \cos\beta$~(\ref{intch}). 
The decay width of $H^- \to \widetilde{\nu}_1 + \widetilde{e}$ 
is given as follows:
%%%
\begin{eqnarray}
\Gamma(H^- \to\widetilde{\nu}_1+\widetilde{e})=
\frac{1}{16\pi m_{H^-}}
|g^{\widetilde{\nu}_1 e H^-}|^2
\kappa^{1/2}\left(
1,\left(\frac{m_{\widetilde{\nu}_1}}{m_{H^-}}\right)^2,
\left(\frac{m_{\widetilde{e}}}{m_{H-}}\right)^2
\right),
\end{eqnarray}
where a function $\kappa$ is defined by
\begin{eqnarray}
\kappa(a,b,c) &\equiv& a^2+b^2+c^2-2ab-2ac-2bc.
\end{eqnarray}
%%% 

\begin{figure}[t]
\begin{center}
\includegraphics[width=6.5cm,clip]{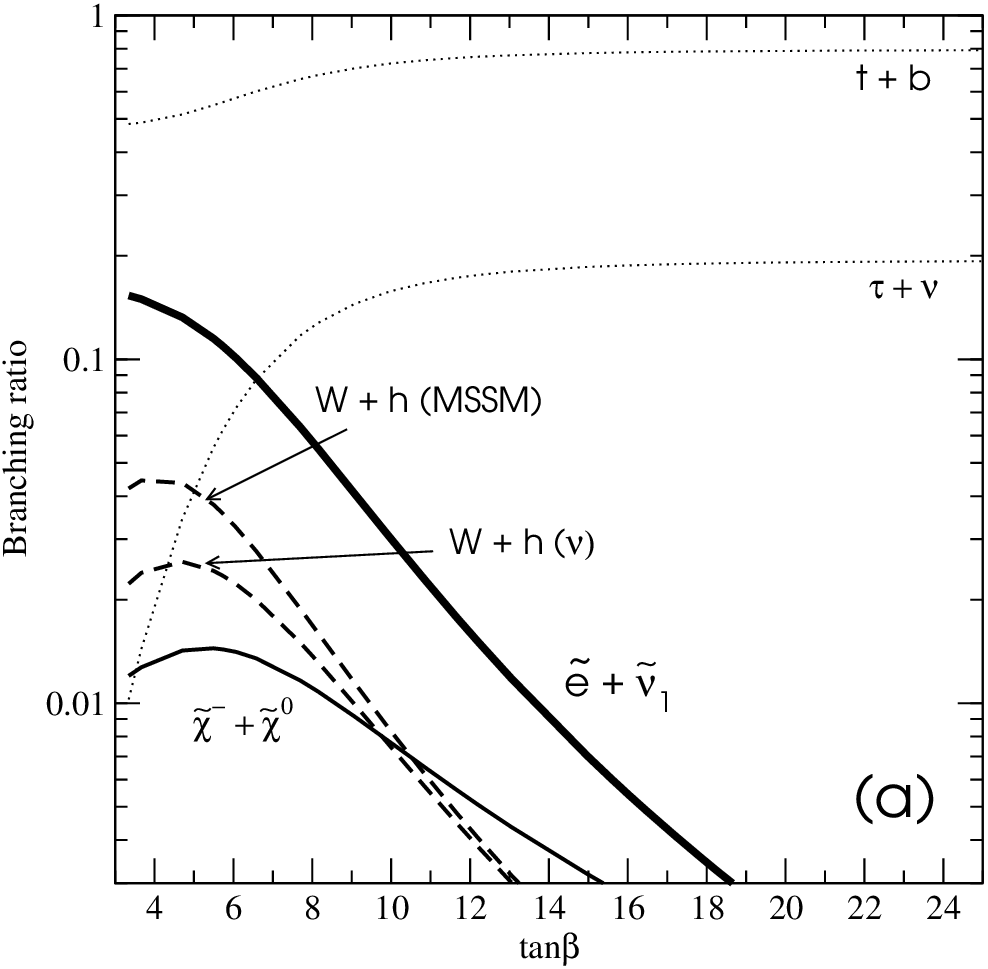}
\includegraphics[width=6.5cm,clip]{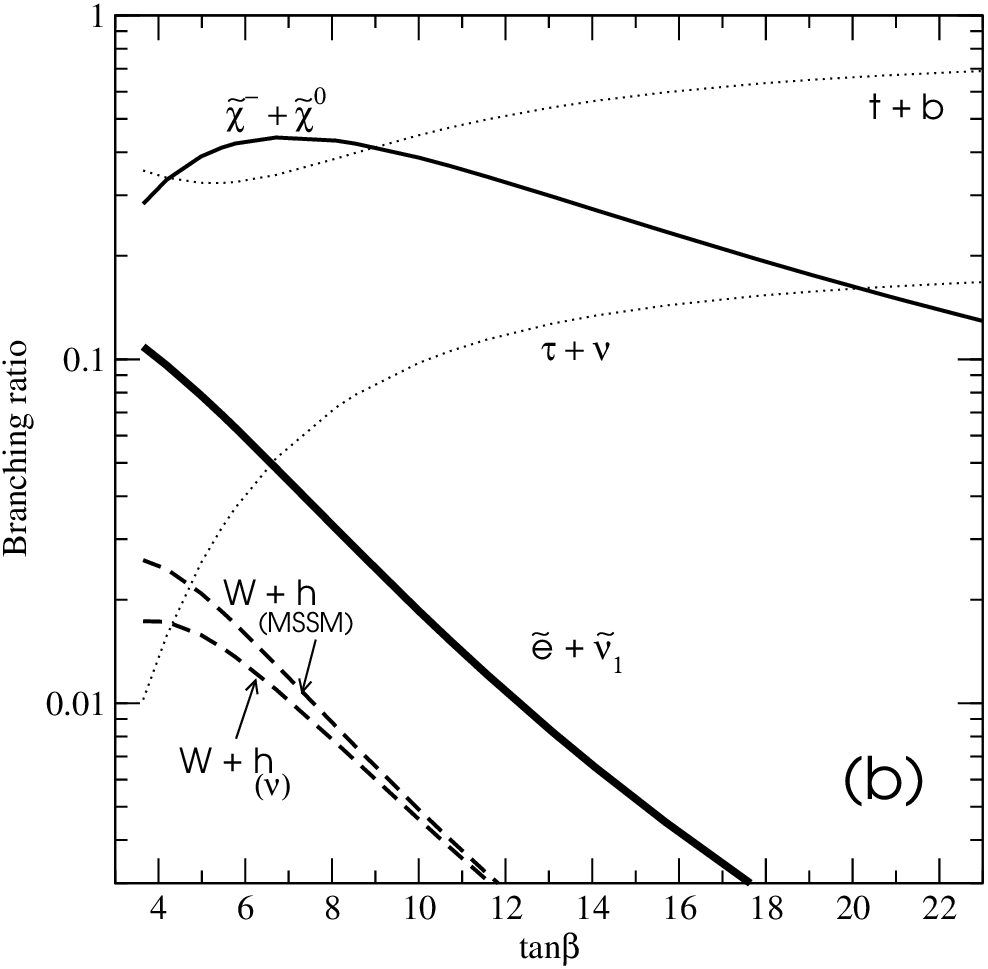}
\end{center}
\caption{The branching ratios of charged Higgs boson decay for
$m_{H^-}=350{\rm GeV}$. 
The decay mode into sneutrino and selectron is found for 
$m_{\tilde{e}_L}=200{\rm GeV}, m_{\tilde{\nu}_1}=50{\rm GeV}, 
A_{\nu}=500{\rm GeV}$. 
The chargino-neutralino mode is obtained for 
$m_{\widetilde{\chi}^-_1}=150{\rm GeV}$ with 
$M_2/\mu=5$ (a) and 1 (b). 
}
\label{brs}
\end{figure}

%%%
In Fig.~\ref{brs}, we show branching ratios of some decay 
modes of the charged Higgs boson with $m_{H^-}=350{\rm GeV}$ 
as functions of $\tan\beta$. 
We assume that squarks are heavy enough so that the decay modes 
into squarks are kinematically forbidden. Heavy squarks are also 
favored to make the lightest Higgs boson heavy through 
the radiative corrections, against for the negative contribution 
to $m_h$ from the sneutrino exchanging diagrams. 
The sneutrino and selectron masses are chosen as 
$m_{\widetilde{\nu}_1}=50{\rm GeV}$ and 
$m_{\widetilde{e}_L}=200{\rm GeV}$, respectively. 
The trilinear coupling of right-handed sneutrino $A_\nu$ 
is fixed at $500{\rm GeV}$. 
Then the heavier sneutrino mass ($m_{\widetilde{\nu}_2}$) is 
about $700{\rm GeV}$. 
As already mentioned, we assumed the flavor universality of 
$A_\nu$, so the branching ratio of decay into the sneutrino 
and smuon, or stau, is same with the selectron mode shown in 
the figure. 
As an example, the branching ratio of decay into charginos 
($\widetilde{\chi}^-_i,i=1,2$)  and neutralinos 
($\widetilde{\chi}^0_j,j=1,4$) is examined for 
$m_{\widetilde{\chi}^-_1}=150{\rm GeV}$ with $M_2/\mu=5$ 
in Fig.~\ref{brs}(a) and $M_2/\mu=1$ in Fig.~\ref{brs}(b), 
where $M_2$ and $\mu$ stand for the SU(2)$_L$ gaugino mass 
and the higgsino mass, respectively. 
The U(1)$_Y$ gaugino mass $M_1$ is obtained using the 
GUT relation, $M_1/\alpha_Y=(5/3)(M_2/\alpha_2)$, where 
$\alpha_i(i=Y,2)$ are given as $\alpha_i=g^2_i/(4\pi)$. 
Then the mass of lightest neutralino is given as 
$m_{\widetilde{\chi}^0_1}\sim 142{\rm GeV}$ in Fig.~\ref{brs}(a) 
and $93{\rm GeV}$ in Fig.~\ref{brs}(b). 
The ratio $M_2/\mu$ determines the properties of the 
lighter chargino and the lightest neutralino. 
When $M_2/\mu \ll 1$ the lighter chargino is mostly the SU(2)$_L$ 
gaugino while the relation $M_2/\mu \gg 1$ corresponds to the 
higgsino dominant case. 
For $M_2/\mu=5$, both the lighter chargino and the lightest 
neutralino are higgsino dominant, so that the decay 
$H^- \to \widetilde{\chi}^-_1+ \widetilde{\chi}^0_1$ 
is highly suppressed because there is no Higgs-higgsino-higgsino 
coupling. 
This explains the difference of 
${\rm Br}(H^- \to \widetilde{\chi}^-_1+ \widetilde{\chi}^0_1)$ 
between Figs.~\ref{brs}(a) and (b). 
%%%

%%%
It can be seen from Fig.~\ref{brs} that the branching ratio of 
$H^- \to \widetilde{\nu} + \widetilde{\ell}$ mode could be 
as large as $10\%$ for small $\tan\beta(\lsim 7)$. 
In the MSSM, the charged Higgs boson can decay into  
$\widetilde{\nu}_L$ and $\widetilde{e}_R$. 
For comparison, we fix the mass of $\widetilde{e}_R$ as 
$m_{\widetilde{e}_R}=m_{\widetilde{e}_L}=200{\rm GeV}$. 
Then the decay mode 
$H^- \to \widetilde{\nu}_L + \widetilde{e}_R$ is kinematically 
forbidden because the sneutrino $\widetilde{\nu}_L$ cannot be 
much lighter than $\widetilde{e}_L$ 
due to the SU(2)$_L$ relation (\ref{mass}) 
(note that $m_{\widetilde{e}_R}=m_{\widetilde{e}_L}=200{\rm GeV}$). 
Therefore, if the charged Higgs boson mass does not differ 
so much from the masses of charged sleptons, the decay 
$H^- \to \widetilde{\nu}_L + \widetilde{e}_R$ in the MSSM 
is strongly suppressed. 
%%%

%%%
Next we study a signal of the decay 
$H^- \to \widetilde{\nu}_1 + \widetilde{e}_L$ in some detail. 
For our choice of the inputs used in Fig.~\ref{brs}, 
the selectron $\widetilde{e}_L$ dominantly decays into the 
lightest neutralino and an electron, 
$\widetilde{e}_L \to \widetilde{\chi}^0_1 + e$. 
Then, since the branching ratio of the 
$\widetilde{\nu}_1+\widetilde{e}_L$ mode is roughly 10\% 
for small $\tan\beta$ region, a probability which we find an 
electron from this decay mode can be estimated as 
${\rm Br}(H^- \to \widetilde{\nu}_1+\widetilde{e}_L) 
\times {\rm Br}(\widetilde{e}_L \to e + \widetilde{\chi}^0_1) 
\simeq 10\%$. 
The electron is also coming out from the $W$ boson of the decay 
$H^-\to W+h$, and the chargino of the decay 
$H^- \to \widetilde{\chi}^- + \widetilde{\chi}^0$. 
From Fig.~\ref{brs} we find that 
${\rm Br}(H^-\to W+h) \lsim 3\%$ and the leptonic decay of 
the $W$ boson is known as 
${\rm Br}(W \to \nu +e ) \lsim 10.8\%$~\cite{Yao:2006px}. 
It leads to 
${\rm Br}(H^-\to W+h) \times {\rm Br}(W \to \nu +e ) \lsim 0.3\%$.  
In case of Fig.~\ref{brs}(a), therefore, the background from 
$H^-\to W+h$ is much suppressed. 
In case of $H^- \to \widetilde{\chi}^- + \widetilde{\chi}^0$, 
the branching ratio is 
${\rm Br}(H^- \to \widetilde{\chi}^- + \widetilde{\chi}^0)$ is 
about $1\%$ and 
${\rm Br}(\widetilde{\chi}^- \to e+\widetilde{\nu})$ is roughly 
$30\%$ per each lepton flavor. 
Thus 
${\rm Br}(H^- \to \widetilde{\chi}^- + \widetilde{\chi}^0)\times 
{\rm Br}(\widetilde{\chi}^- \to e+\widetilde{\nu})$ is about $0.3\%$. 

As shown in Fig.~\ref{brs}(b), however, if the lighter chargino is 
dominantly gaugino, the branching ratio of the chargino-neutralino 
mode increases, so that the branching ratio of 
$H^- \to \widetilde{\nu}_1 + \widetilde{e}_L$ is relatively 
decreased. 
In this case we estimate 
the probability that the electron is found in the 
$\widetilde{\chi}^- + \widetilde{\chi}^0$ mode of the 
charged Higgs decay as 
${\rm Br}(H^- \to \widetilde{\chi}^- + \widetilde{\chi}^0)
\times 
{\rm Br}(\widetilde{\chi}^- \to e + \widetilde{\nu})
\simeq 10\%$. 
This competes with the probability that an electron is coming 
out from the $\widetilde{e}_L + \widetilde{\nu}_1$ decay. 
We conclude that, even in our specific choice of parameter 
set, the $\widetilde{\chi}^- + \widetilde{\chi}^0$ mode could 
be a serious background to search the decay 
$H^- \to \widetilde{\nu}_1 + \widetilde{e}_L$ when the chargino 
and neutralino are almost gauginos. 
%%% 
%\vspace{1cm}

%%%-----------------------------
%%% Fig. 3
%%%-----------------------------
%%%%%%% 

We would like to discuss the testability of the scenario of 
light $\widetilde{\nu}_R$ with unsuppressed $A_\nu$ 
at future collider experiments using the decay 
$H^- \to \widetilde{\nu}_1 + \widetilde{e}_L \to e + \MEt$. 
An important point is to identify that the observed 
electron comes from $H^-$. 
It could be achieved using the pair production of the charged Higgs 
bosons. 
In a pair production of the charged Higgs, one of the charged Higgs 
bosons can be identified using the $t+b$ mode. Then if an electron 
is observed in the charged Higgs pair production it must be identified  
as one from the decay of another charged Higgs through 
$H^- \to \widetilde{\nu}_1 + \widetilde{e}_L$. 
For example, at the $e^+ e^-$ linear collider (ILC), 
the typical size of the cross section of the charged Higgs boson 
pair is ${O(1-10)}$(fb) for $m_{H^-}=O(100{\rm GeV})$~\cite{higgs}. 
Assuming the integrated luminosity as $100{\rm fb}^{-1}$, it is expected 
that $100\sim 1000$ charged Higgs pairs are produced in a year. 
Fig. \ref{brs}(a) tells us that, when $\tan\beta=3$, 
only few electrons appear from $1000$ charged Higgs bosons in 
the MSSM (the $W+h$ mode), while 
about $160$ electrons from the $\widetilde{e}+\widetilde{\nu}_1$ 
mode is expected in our scenario. 
Therefore, an excess of electrons from the charged Higgs decay could 
be a signal of the TeV scale right-handed sneutrino with unsuppressed 
trilinear coupling $A_\nu$. 

%%%%%%%%%%%%%%%%%%%%%%%
%%%%%%%%%%%%%%%%%%%%%%%
\section{Summary}
\label{sum}
%%%%%%%%%%%%%%%%%%%%%%%
%%%%%%%%%%%%%%%%%%%%%%% 

In this paper, we have studied phenomenology of the scenario of 
TeV scale right-handed sneutrino inspired by models of SUSY 
breaking inspired neutrino 
mass~\cite{Mur,Borzumati:2000mc,March-Russell:2004uf}.  
The important prediction of this scenario is that the sneutrino 
trilinear coupling $A_{\nu}$ could be sizable and is not suppressed  
by the neutrino Yukawa coupling. 
We examined two phenomenological consequences of this scenario. 
We found that the sneutrino contribution to the lightest Higgs 
boson mass is destructively interferes with the ordinary MSSM
contributions. 
Thus the lightest Higgs boson mass may be lowered in this 
model via sneutrino exchange with large $A_{\nu}$. 
The large $A_{\nu}$ also affects the decay of 
charged Higgs boson. 
It is shown that the process 
$H^- \to \widetilde{\nu}_1+\widetilde{e}_L$ could be subdominant 
decay mode in some parameter region and the branching ratio 
is roughly $\sim 10\%$ for small $\tan\beta$. 
In such parameter region, we expect that roughly 200 electrons 
per year from the charged Higgs decay at the ILC experiments 
with the integrated luminosity $100{\rm fb}^{-1}$. 
On the other hand the MSSM predicts only few electrons from the 
charged Higgs decay. The excess of the electrons in the charged 
Higgs decay, therefore, could be a signal of the TeV 
$\widetilde{\nu}_R$ scenario. 

\vspace{6mm}
\section*{Acknowledgments}
The work of G.C.C. was supported in part by the Grant-in-Aid for 
Science Research, Ministry of Education, Science and Culture, 
Japan (No.K175402386). 
The work of S.K. was supported by the Japan Society of Promotion 
of Science. 

%%%%%%%%%%%%%%%%%%%%%%%
%%%%%%%%%%%%%%%%%%%%%%%
%%%%%%%%%%%%%%%%%%%%%%%

\end{document}